\documentclass[a4paper,11pt]{article}
\usepackage{graphicx}
\usepackage{tabularx}
\usepackage{multicol}
\usepackage{hyphenat}
\usepackage{amsmath,amssymb,amsfonts,amsthm,latexsym}
\usepackage{pos}
\usepackage{slashed}
\usepackage{units}
\usepackage{siunitx}
\usepackage{subcaption}
\usepackage{lineno}

\newcommand{\ba}{\begin{linenomath}\begin{align}} 
\newcommand{\ea}{\end{align}\end{linenomath}}

\newcommand{\lb}{\left(}
\newcommand{\rb}{\right)}

\newcommand{\lsb}{\left[}
\newcommand{\rsb}{\right]}
\newcommand{\bra}{\left\langle}
\newcommand{\ket}{\right\rangle}

\title{Implications of gradient flow on the static force}
\ShortTitle{Static force with gradient flow}

\author*[a,b]{Julian Mayer-Steudte}
\author[a,b,c]{Nora Brambilla}
\author[d]{Viljami Leino}
\author[a]{Antonio Vairo}

\affiliation[a]{Physik Department, Technische Universit\"at M\"unchen,
James-Franck-Strasse 1, 85748 Garching, Germany}
\affiliation[b]{Munich Data Science Institute, Technische Universit\"at M\"unchen, \\
Walther-von-Dyck-Strasse 10, 85748 Garching, Germany}
\affiliation[c]{Institute for Advanced Study, Technische Universit\"at M\"unchen,
Lichtenbergstrasse 2a, 85748 Garching, Germany}
\affiliation[d]{Helmholtz Institut Mainz, Johannes Gutenberg-Universität Mainz, 55099 Mainz, Germany}

\emailAdd{julian.mayer-steudte@tum.de}
\emailAdd{nora.brambilla@ph.tum.de}
\emailAdd{viljami.leino@tum.de}
\emailAdd{antonio.vairo@ph.tum.de}

\abstract{We use gradient flow to compute the static force based on a Wilson loop with a chromoelectric field insertion.
The result can be compared on one hand to 
the static force from the numerical derivative of the lattice static energy, and on the other hand to the perturbative calculation, allowing 
a precise extraction of the    $\Lambda_0$ parameter. 
This  study may open the way to gradient flow calculations of 
correlators of chromoelectric and chromomagnetic fields,
which typically arise in the nonrelativistic effective field theory factorization.}

\FullConference{%
The 39th International Symposium on Lattice Field Theory,\\
8th-13th August, 2022,\\
Rheinische Friedrich-Wilhelms-Universität Bonn, Bonn, Germany
}

\begin{document}
\maketitle

\section{Introduction}

The static energy $E(r)$ of a quark-antiquark pair is one of the first quantities computed in lattice QCD; the static force is its derivative. At short distances it can be compared to perturbative calculations, and therefore, it can be used for extracting the strong coupling constant. Furthermore, it can be used to set the scale on the lattice.

The static energy is perturbatively known up to N$^3$LL~\cite{Brambilla:1999qa,Pineda:2000gza,Brambilla:2006wp,Anzai:2009tm,Smirnov:2009fh} and, hence, so is the force $F(r)$. However, the perturbative static energy in dimensional regularization 
has a renormalon ambiguity of order $\Lambda_\mathrm{QCD}$, which corresponds to a linear UV divergence ($\propto 1/a$) on the lattice. This ambiguity and divergence vanish if we take the derivative for obtaining the static force, and all the
physical information, like the coupling, is encoded  
in the slope of the static energy.

Recently, the force has been measured on the lattice directly~\cite{Brambilla:2021wqs}, and perturbatively 
calculated at NLO in continuum in $\overline{\textrm{MS}}$  with gradient flow~\cite{Brambilla:2021egm}. 
The present work is based on both developments 
and presents updates with respect to  previous  results~\cite{Leino:2021vop}.  We find it useful to use 
the  derivative of the static energy as a benchmark
to assess  discretization effects and systematics.
This work is also instrumental to set best strategies to calculate on the lattice chromoelectric and chromomagnetic field correlators arising in the nonrelativistic effective field 
theories low energy factorization.
The study is sufficiently innovative to grant interest even 
if quenched.

The proceedings are organized as follows.
In Sec.~\ref{sec:physical_setup} we introduce 
the definition of the force in terms of a chromolectric insertion in a static Wilson loop, the gradient flow, the flowed
 force and the   lattice setup. In Sec.~\ref{sec:analysis}, we present the analysis of the lattice data, the continuum limit and the comparison to the perturbative results in the short range.

\section{Physical setup}\label{sec:physical_setup}

\subsection{The static force}\label{subsec:static_force}

The static energy $E(r)$ is related to a Wilson loop $W_{r\times T}$ with temporal extent from $0$ to $T$ and spatial extent  $r$~\cite{Wilson:1974sk} by
\begin{linenomath}\begin{align}
    E(r) &= -\lim_{T\rightarrow \infty} \frac{\ln\langle \mathrm{Tr}(W_{r\times T})\rangle}{T} = - \frac{1}{a}\lim_{T\rightarrow \infty} \frac{\langle \mathrm{Tr}(W_{r\times(T+a)})\rangle}{\langle \mathrm{Tr}(W_{r\times T})\rangle}\,,\\
    W_{r\times T} &= P\left\{ \exp\left( i\oint _{r\times T} dz_\mu gA_\mu \right) \right\}\,,
\end{align}\end{linenomath}
where $a$ is the lattice spacing, and $P$ is the path ordering operator. 
The static force $F(r)$ is defined as 
the derivative of the static energy:
\begin{linenomath}\begin{align}
    F(r) = \partial_r E(r).
\end{align}\end{linenomath}
Often, this derivative is evaluated from the static energy data either with interpolations or with
finite differences. However, this leads to increased systematic errors.
Alternatively, 
we can measure the force on the lattice directly via 
the formula ~\cite{Vairo:2015vgb,Vairo:2016pxb,Brambilla:2000gk}
\begin{linenomath}\begin{align}
    F(r) &= -\lim _{T\rightarrow\infty}\frac{i}{\langle \mathrm{Tr}(W_{r\times T} \rangle}\left\langle\mathrm{Tr}\left( P\left\{ \exp\left( i\oint _{r\times T} dz_\mu gA_\mu \right) \mathbf{\hat{r}}\cdot g\mathbf{E}(\mathbf{r}, t^*) \right\} \right) \right\rangle \\
    &= -\lim _{T\rightarrow\infty} i \frac{\langle \mathrm{Tr}\{ PW_{r\times T}gE_j(\mathbf{r},t^* \} \rangle}{\langle \mathrm{Tr}(W_{r\times T} \rangle}\,,
\end{align}\end{linenomath}
where the expression in the numerator consists of a static  Wilson loop with a chromoelectric field insertion on the temporal Wilson line at position $t^*$, and $\mathbf{\hat{r}}$ is the spatial direction. In general, $t^*$ is arbitrary. Nevertheless, we choose $t^*=T/2$ for even-spaced separations, and an average over $t^*=T/2\pm a/2$ for odd-spaced separations. This reduces the interactions between the $E$-field and the corners of the Wilson loop.

\subsection{Gradient flow}\label{subsec:gradient_flow}
The Yang-Mills gradient flow flows the gauge fields $U_\mu$ towards the minimum of the Yang-Mills gauge action along a fictitious dimension called flow time $\tau_F$. The transformation is determined by solving the ordinary differential equation for flowed link variables with the original configuration as initial condition~\cite{Narayanan:2006rf,Luscher:2009eq,Luscher:2010iy}:
\begin{linenomath}\begin{align}
    \Dot{V}_{\tau_F}(x,\mu) &= -g_0^2\lsb \partial_{x,\mu}S(V_{\tau_F}) \rsb V_{\tau_F}(x,\mu)\,,\\
    V_{\tau_F}(x,\mu)|_{\tau_F=0} &= U_\mu(x)\,,
\end{align}\end{linenomath}
where $\partial_{x,\mu}S(V_{\tau_F})$ is the derivative of a gauge action according to the flowed link variable $V_{\tau_F}(x,\mu)$. It is known that gradient flow acts like smearing characterized by the flow radius $\sqrt{8\tau_F}$, which improves the signal to noise ratio and renormalizes 
gauge invariant operators. This is especially useful for operators with field strength component insertions, which would normally require an extra renormalization on the lattice.

The flowed link variable depends only on the initial configuration, therefore, we can write it as a function of the non-flowed field as $V_{\tau_F}=V_{\tau_F}[U]$. Flowed observables are obtained by replacing the original gauge links with the flowed ones, the resulting path integral is represented as
\begin{linenomath}
    \begin{align}
        \bra O(\tau_F) \ket = \frac{1}{Z}\int \mathcal{D}[U] e^{-S_E[U]}O[V_{\tau_F}]\,,
    \end{align}
\end{linenomath}
which states that the underlying theory remains unchanged. However, to connect to the physical observable, we have to perform the limit $\tau_F\rightarrow 0$.

\subsection{The perturbative force in gradient flow}\label{subsec:flowed_force}

The static force in gradient flow has been perturbatively determined at 1-loop order in Ref.~\cite{Brambilla:2021egm} for general flow time $\tau_F$. In the small flow time expansion, it can be expressed as
\begin{linenomath}
    \begin{align}
        r^2F^\mathrm{1-loop}(r,\tau_F) \approx r^2F^\mathrm{1-loop}(r,\tau_F=0) + \frac{\alpha_S^2C_F}{2\pi}\underbrace{\lsb -12\beta_0-6C_Ac_L\rsb }_{8n_f}\frac{\tau_F}{r^2}\,,\label{eq:pert_small_flow time_expansion}
    \end{align}
\end{linenomath}
with $c_L=-22/3$. 
We see that, in pure gauge, the static force should be constant for small flow times, since the flow time dependent term is proportional to $n_f$. Furthermore, the relevant scale for the force
is the dimensionless ratio $\tau_F/r^2$. We note that in the analysis, the flow time $\tau_F$ and 
the flow time ratio $\tau_F/r^2$ can be used interchangeably.
The whole 1-loop expression can be found in~\cite{Brambilla:2021egm}. 

Since the static force is known to 3-loop order at zero flow time and the higher loop contributions are crucial for the extraction of the $\Lambda_0$ parameter, we choose a hybrid ansatz, where we model the flow time dependence with the full 1-loop formula and use the higher loop result at zero flow time.
Up to a given order:
\begin{linenomath}
    \begin{align}
        r^2F^\mathrm{order}(r,\tau_F) &\equiv r^2F^\mathrm{order}(r,\tau_F=0) + f^\mathrm{1-loop}(r,\tau_F)\label{eq:pert_force_combined}\\
        f^\mathrm{1-loop}(r,\tau_F) &\equiv r^2F^\mathrm{1-loop}(r,\tau_F) - r^2F^\mathrm{1-loop}(r,\tau_F=0)\,.
    \end{align}
\end{linenomath}

We work in the pure $SU(3)$ gauge theory
($n_f=0$), and  call 
 $\Lambda_0 \equiv \Lambda_{\overline{\textrm{MS}}}^{n_f=0}$. 
 In Ref~\cite{Brambilla:2021egm}, they discuss the scale choices of $\mu=1/r$, $\mu=1/\sqrt{r^2+8\tau_F}$, and $\mu=1/\sqrt{8\tau_F}$. We will focus on $\mu=1/r$ and $\mu=1/\sqrt{r^2+8\tau_F}$ here.

\subsection{Lattice setup}\label{subsec:lattice_setup}

\begin{table}
    \centering
    \begin{tabular}{c|c|c|c|c|c}
        $N_S$ & $N_T$ & $\beta$ & $a[\si{fm}]$ & $N_\mathrm{conf}$ & Label\\\hline
        20 & 40 & 6.284 & 0.060 & 6000 & L20 \\
        26 & 52 & 6.481 & 0.046 & 6000 & L26 \\
        30 & 60 & 6.594 & 0.040 & 6000 & L30 \\
        40 & 80 & 6.816 & 0.030 & 2700 & L40
    \end{tabular}
    \caption{The lattice parameters for the computations.}
    \label{tab:lattice_parameters}
\end{table}

We generate quenched gauge field configurations with the MILC code. Table~\ref{tab:lattice_parameters} shows the parameters used in our calculations. The scale setting is done through \cite{Necco:2001xg}.
The configurations are produced with overrelaxation and heatbath algorithm.
For solving the gradient flow equation, we use the Symanzik improved action, and fixed stepsize algorithm \cite{Luscher:2010iy} for the lattices L20, L26, and L30, and adaptive algorithm for 
L40~\cite{Fritzsch:2013je,Bazavov:2021pik}. We use blocking with jackknife sampling for the error propagation.

\section{Analysis}\label{sec:analysis}

\subsection{Plateau extraction}\label{sec:plateau_extraction}
In order to extract the ground state of the static force and energy 
in the $T\rightarrow \infty$ limit, we fit a constant within a suitable range of $T$. This has to be done for each fixed $(r,\tau_F)$ or $(r,\tau_F/r^2)$ at each lattice. We automatize the plateau extraction with the Akaike information criterion (AIC) based 
model averaging, as described in Ref.~\cite{Jay:2020jkz}. 

Here, we will briefly summarize the model averaging procedure.
We are interested in the model parameters $\mathbf{a}\in \mathbb{R}^k$ ($k$ being the number of parameters) of a model function $f(x,\mathbf{a})$ for a quantity $x_i$ with $i=1, \dots , N$. We have the data $D=(x_i,y_i,C_{ij})$ with $y_i$ being the data corresponding to $x_i$, and $C_{ij}$ its covariance matrix. We perform the fits using the data only in the range $i_1$ to $i_2$ for all possible ranges between $1, \dots , N$, and $i_2-i_1>3$ fixed. Every fit has a set of optimal parameters $\mathbf{a^\star}_{i_1,i_2}$ and an optimal $\chi^2$-value which leads us to the information criterion
\begin{linenomath}
    \begin{align}
        \mathrm{AIC}_{i_1,i_2}=\sum_{i,j=i_1}^{i_2} \lsb f(x_i,\mathbf{a^\star}_{i_1,i_2})-y_i\rsb C^{-1}_{ij} \lsb f(x_j,\mathbf{a^\star}_{i_1,i_2})-y_j\rsb + 2k + 2(i_1-i_2)\,,
    \end{align}
\end{linenomath}
which defines the model probability
\begin{linenomath}
    \begin{align}
        p(i_1,i_2|D) \propto e^{-\frac{1}{2}\mathrm{AIC}_{i_1,i_2}}\,.
    \end{align}
\end{linenomath}
Finally, we are interested in a specific component of the parameter vector, and we can determine the expectation value and the standard deviation as
\begin{linenomath}
    \begin{align}
        \overline{a}_n &= \bra a_n\ket = \sum_{i_1,i_2} a_{n,i_1,i_2}^\mathbf{\star} p(i_1,i_2|D),\\
        \sigma ^2_n &= \bra \lb a_n - \overline{a}_n \rb ^2\ket = \bra a_n^2 \ket - \bra a_n\ket ^2\,.\label{eq:AIC_error}
    \end{align}
\end{linenomath}
For the specific case of constant fit, we have one parameter ($k=1$), the constant $c$, the support points of $T$ correspond to $x_i$, and the model function $f$ is the constant function $f(T,c)=c$.

For our analysis, we measure the AIC expectation value for each jackknife pseudoensemble separately, and then use the jackknife errors for the standard deviation. The error given by the jackknife procedure is comparable to the model uncertainty of Eq.~\eqref{eq:AIC_error}.

\subsection{Impact of the gradient flow}
For the static force, the chromoelectric field inserted in a Wilson loop comes with a lattice only self-energy contribution that needs to be renormalized away in order to attain a proper continuum limit. The renormalization is multiplicative and  independent of the distance $r$:

\begin{linenomath}
    \begin{align}
        F^\mathrm{ren}_\mathrm{latt}(r,\tau_F)=Z_E(\tau_F)F_\mathrm{latt}(r,\tau_F)\,.
    \end{align}
\end{linenomath}
$Z_E$ can be non-perturbatively determined by comparing $F_\mathrm{latt}$ to a numerical derivative of the static energy, which 
does not include field insertions~\cite{Brambilla:2021wqs}:
\begin{linenomath}
    \begin{align}
        Z_E(\tau_F) = \frac{\partial_rV_\mathrm{latt}(r,\tau_F)}{F_\mathrm{latt}(r,\tau_F)}\,.
    \end{align}
\end{linenomath}
We extract $Z_E$ at fixed flow time by performing a constant fit within an intermediate $r$-range,
that is not affected by finite size or finite volume effects. 
We use the AIC model averaging for these fit ranges 
as described in Sec.~\ref{sec:plateau_extraction}.
\begin{figure}
    \centering
    \includegraphics[width=0.47\textwidth]{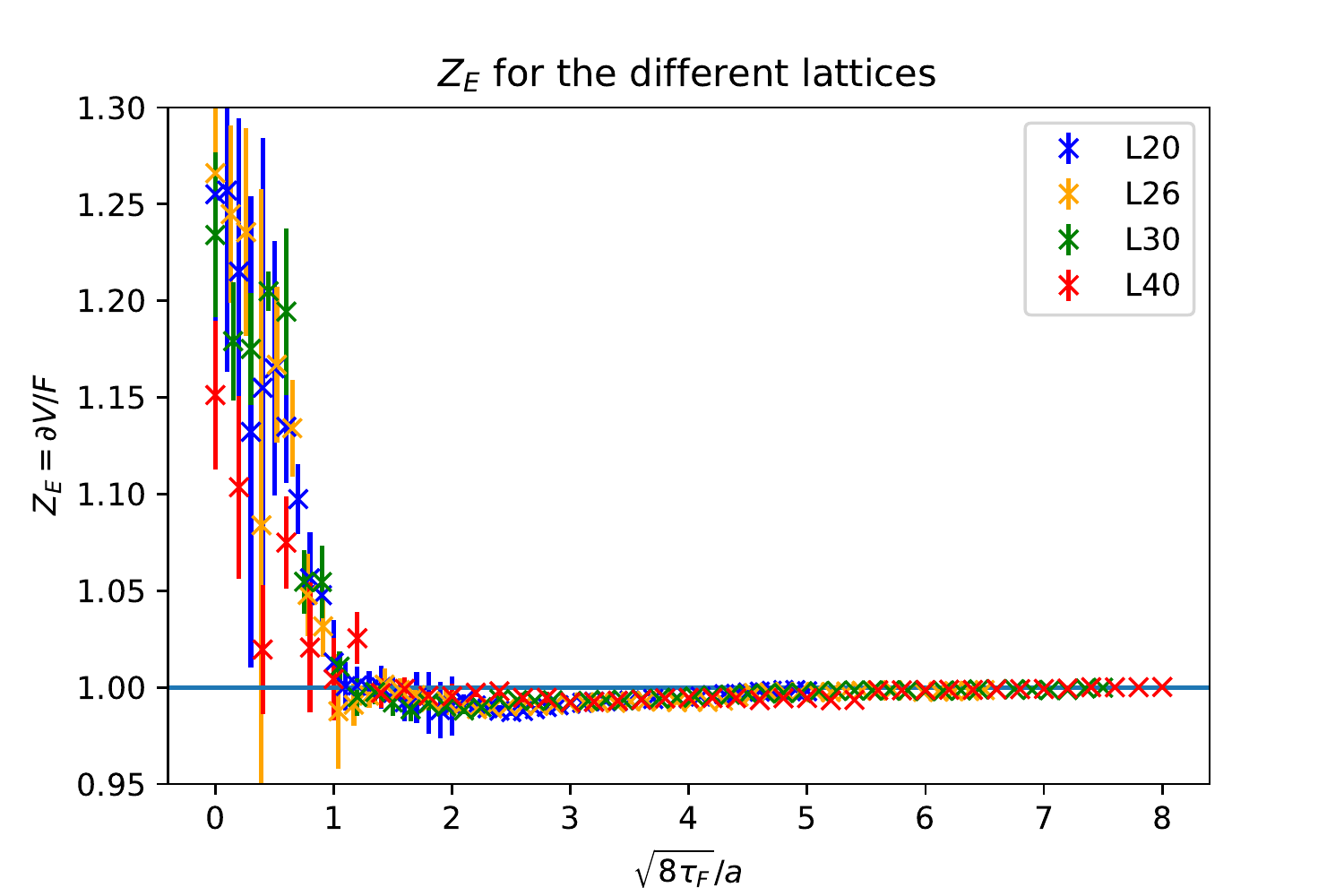}
    \includegraphics[width=0.47\textwidth]{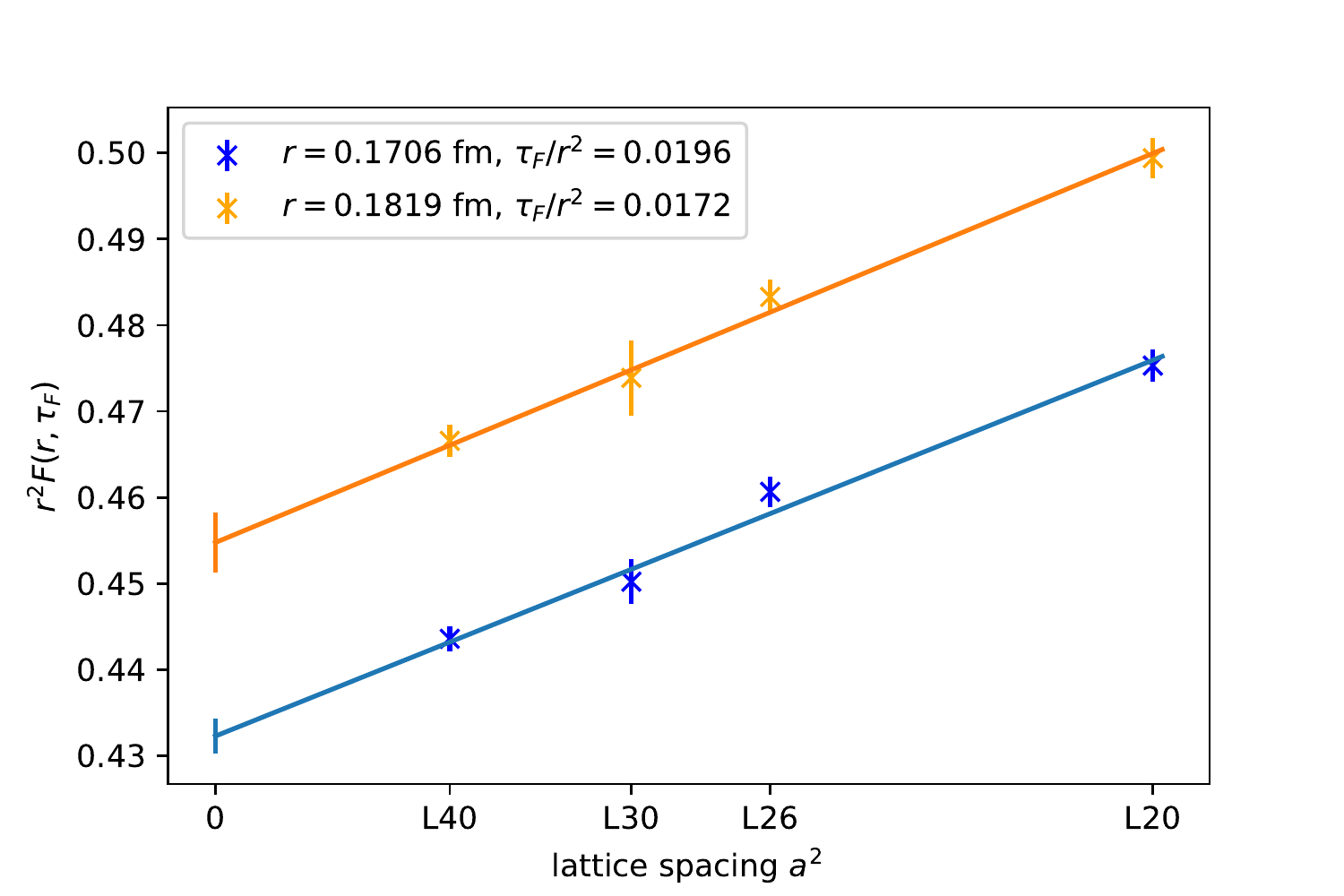}
    \caption{Left: The non-perturbative determination of the flowed renormalization constant $Z_E$. Right: An example for a continuum limit where $Z_E\approx 1$.}
    \label{fig:Ze_flowed}
\end{figure}
The left side of Fig.~\ref{fig:Ze_flowed} shows $Z_E$ for the different lattice sizes, the $x$-axis is in units of the flow radius in lattice units. We see that, for flow radii larger than one lattice spacing ($\sqrt{8\tau_F}\gtrsim a$), $Z_E$ approaches 1, meaning that gradient flow reduces effectively the discretization artifacts. This allows us to perform reliable continuum limits at finite flow time. We consider continuum limit data only for flow times satisfying the condition $\sqrt{8\tau_F}/a\gtrsim 1$. An example for a continuum limit is shown on the right side of Fig.~\ref{fig:Ze_flowed}.

\subsection{Static force at large $r$}

\begin{figure}
    \centering
    \includegraphics[width=0.44\textwidth]{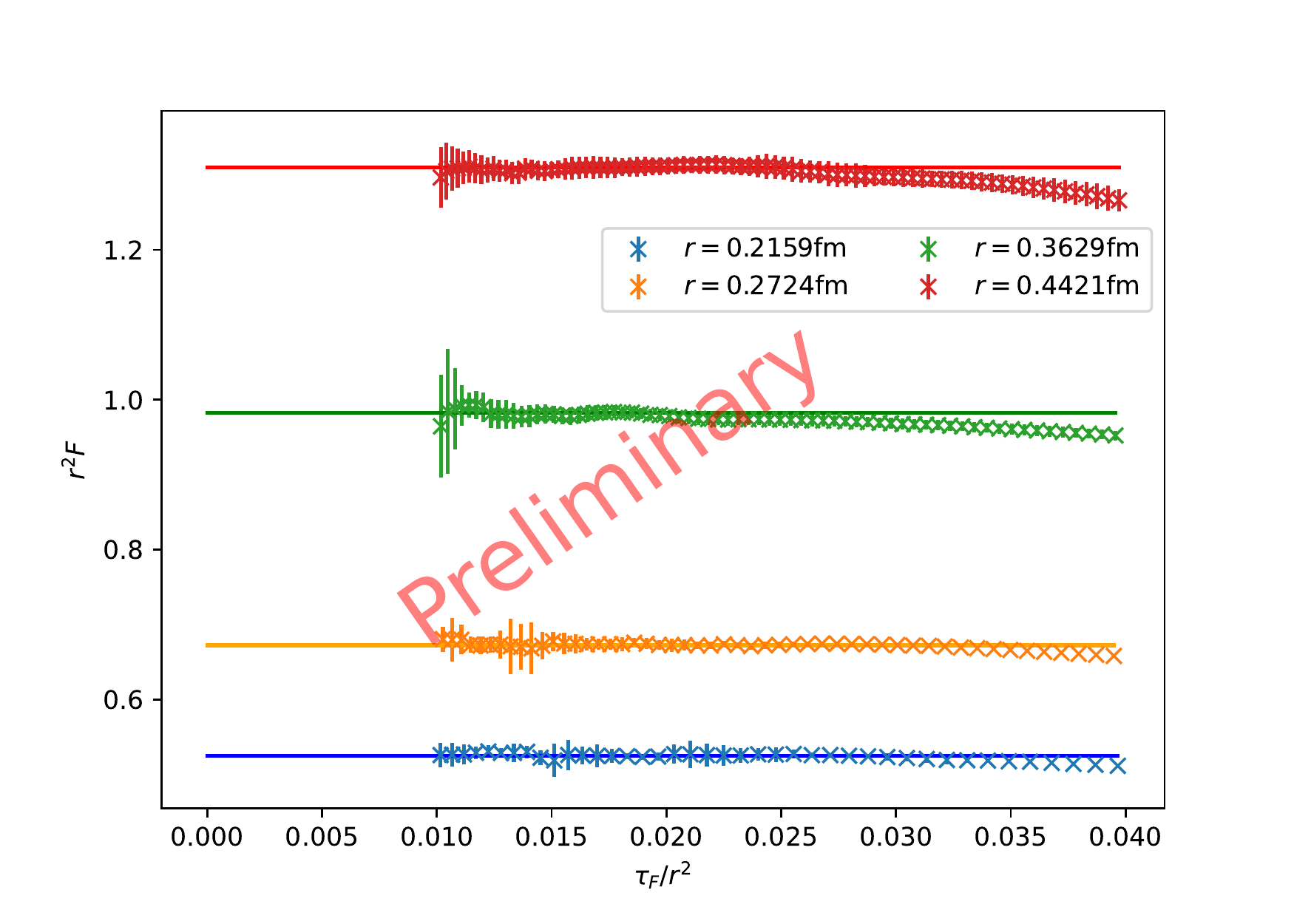}
    \includegraphics[width=0.47\textwidth]{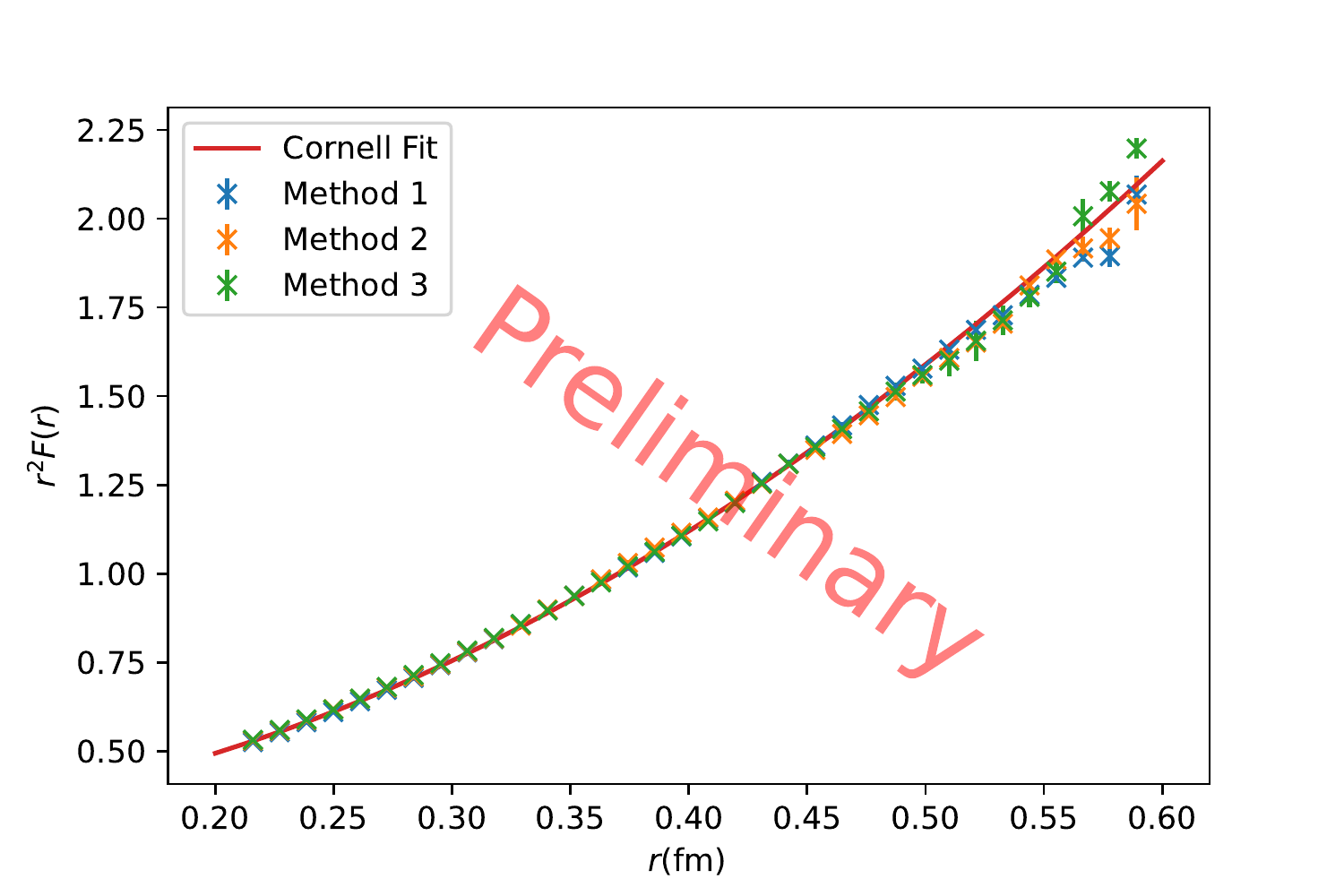}
    \caption{Left: The flowed force at fixed $r$, for large $r$. The $x$-axis is the flow time ratio. Right: The results of the AIC range averaged constant fits.}
    \label{fig:flowed_force}
\end{figure}
According to Eq.~\eqref{eq:pert_small_flow time_expansion}, we expect a constant behavior in the small flow time ratio expansion. The left side of Fig.~\ref{fig:flowed_force} shows the  force at large fixed $r$. We see a constant regime within the range $0.01\leq \tau_F/r^2\leq 0.025$ and we use AIC as described in Sec.~\ref{sec:plateau_extraction} to consider   all meaningful ranges to reduce the systematic errors arising from the choice of the fit range. The results of the constant fit are shown on the right side of Fig.~\ref{fig:flowed_force}. 
The figure shows the results of three different methods: they differ by interpolation functions and whether the continuum limit was performed at fixed flow time or at fixed flow time ratio. All methods should give the same results, however, we see  deviations at larger $r$ indicating the size of the systematic errors. We fit the Cornell ansatz
\begin{linenomath}
    \begin{align}
        r^2F(r) = A + \sigma r^2
    \end{align}
\end{linenomath}
to the result, with $\sigma$ being the string tension. We obtain $\sigma$ between 
$5.18 \si{fm^{-2}}$ and 
$5.23\si{fm^{-2}}$ and $A$ between 
$0.2853$ and $0.2954$. We note that the distances achievable with our lattice sizes are too small for reliable measurement of the string tension.

\subsection{Static force at small $r$}
\begin{figure}
    \centering
    \includegraphics[width=0.47\textwidth]{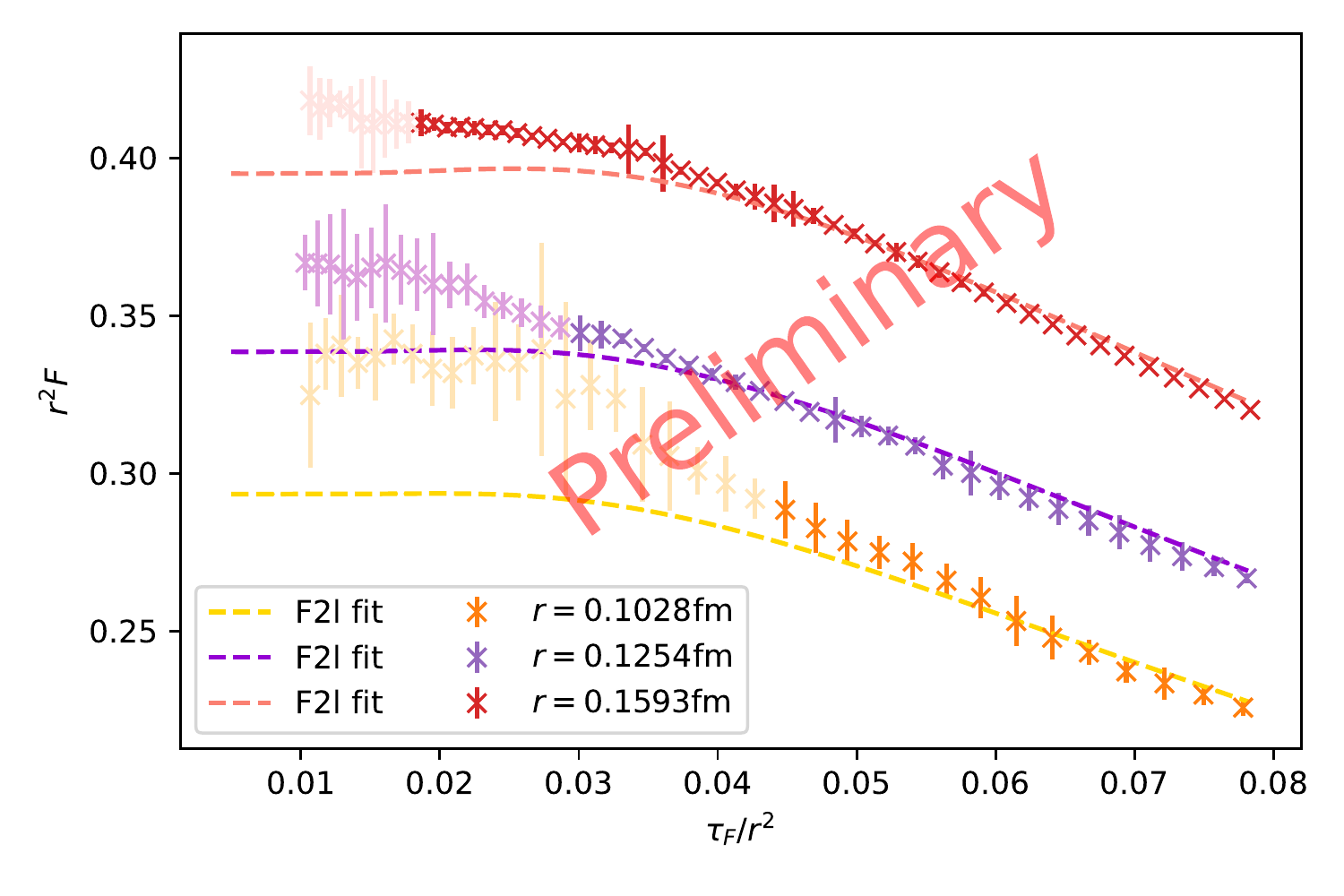}
    \includegraphics[width=0.47\textwidth]{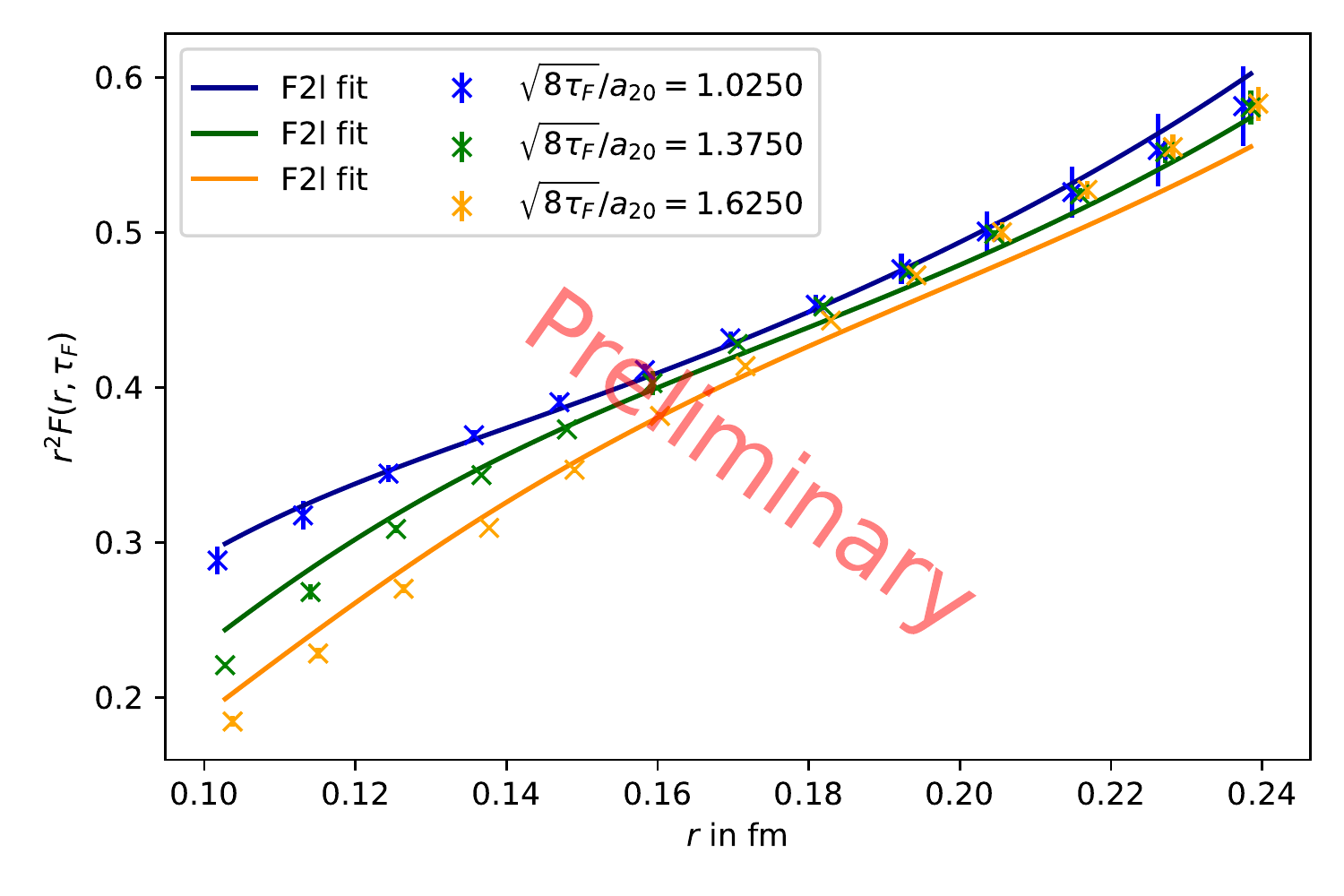}
    \caption{Fitting the perturbative force (dashed lines) to the data (crosses and errorbars) at small $r$. Left:  Fit for fixed $r$, the dimmer points represent the points where not enough flow time was applied and therefore, $Z_E\neq 1$. Those points are excluded from the fits. Right: Fit   for a fixed $r$-range at three different exemplary flow times. In all fits, $\Lambda_0$ is the fit parameter.}
    \label{fig:flowed_force_small_r}
\end{figure}

The left side of Fig.~\ref{fig:flowed_force_small_r} shows the force at fixed small $r$ against the flow time ratio. The dimmer points at smaller flow time ratio belong to the regime where $Z_E\neq 1$. 
We fit Eq.~\eqref{eq:pert_force_combined} treating $\Lambda_0$ as the fit parameter to the data in the valid regime. For the three cases shown on the left side in Fig.~\ref{fig:flowed_force_small_r}, we obtain a range of $\Lambda_0$ between $0.238$ 
and $0.256$ {GeV}. The scale choice $\mu=1/r$ fits best to our measured data in this case.

\begin{figure}
    \centering
    \includegraphics[width=0.47\textwidth]{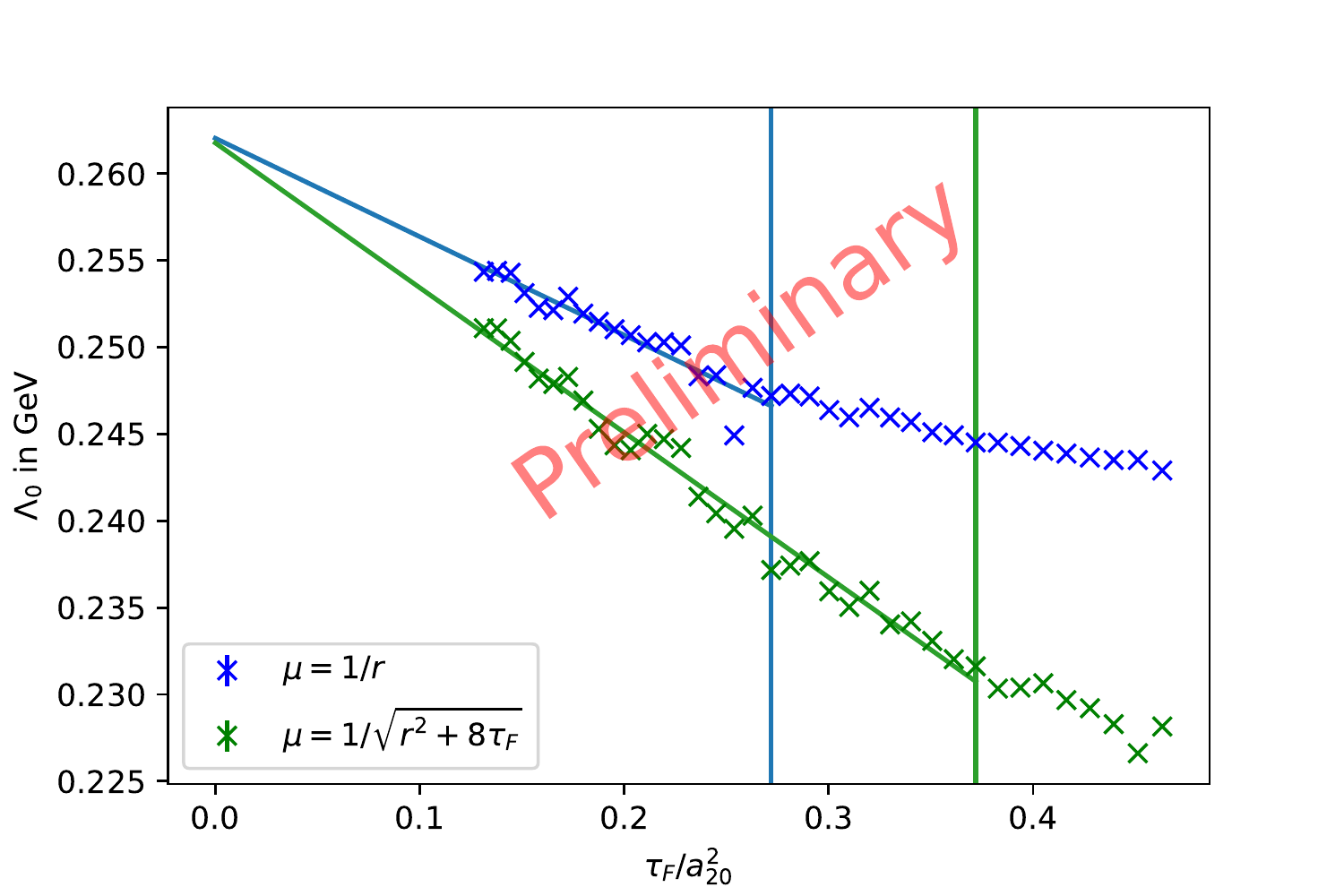}
    \includegraphics[width=0.47\textwidth]{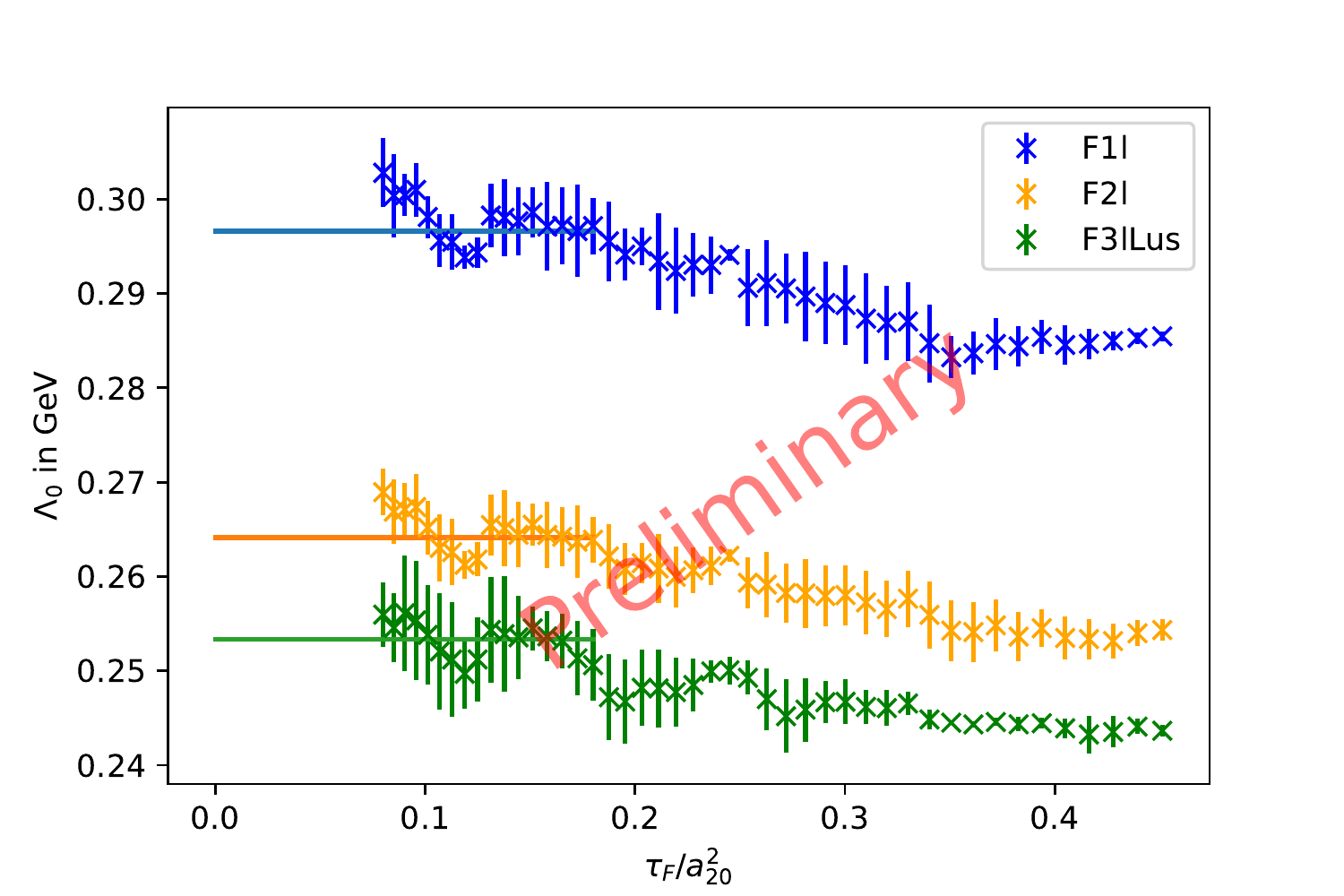}
    \caption{The results for the $\Lambda_0$ fits at fixed $r$ for different flow times (x-axis) in units of the L20 lattice spacings $a_{20}=\SI{0.06}{fm}$. With these units, we can easily show the threshold ($\tau_F/a_{20}^2>0.125$) for including or excluding the L20 points to the continuum limit. Left: The result at fixed $r$-range. Right: The model averaged result for $\Lambda_0$ with averaging over different $r$-ranges. The solid lines show the zero flow time extrapolation.}
    \label{fig:fixed_tau_lambda0}
\end{figure}

Instead of fitting to data at fixed $r$, we can fit to fixed $\tau_F$ along the $x$-axis. It turns out that this procedure works better for fixed flow time than for fixed flow time ratio. The right side of Fig.~\ref{fig:flowed_force_small_r} shows the fit at three different fixed flow times, the left side of Fig.~\ref{fig:fixed_tau_lambda0} shows the resulting $\Lambda_0$ at finite $\tau_F$ for two scale choices. Both choices seem to exhibit a linear flow time behavior in a proper range of flow time. We can assume a linear ansatz within the linear regime and perform 
a zero flow time limit for $\Lambda_0$ which gives $\Lambda_0=\SI{0.262}{GeV}$ and $\Lambda_0=\SI{0.261}{GeV}$ respectively. This is close to the $\Lambda_0$ 
from FLAG~\cite{FlavourLatticeAveragingGroupFLAG:2021npn} with $\Lambda_0^\mathrm{FLAG}=\SI{261(15)}{MeV}$.
We should not see any 
flow time dependence on $\Lambda_0$. 
Therefore, the flow time dependence, we observe in Fig.~\ref{fig:fixed_tau_lambda0}, 
must arise either from corrections at higher orders of perturbation theory or
from systematic effects in the chosen analysis. 

Instead of a fixed $r$-range, we can use the AIC procedure from Sec.~\ref{sec:plateau_extraction}
to find an optimal range for the $\Lambda_0$-fit.
The resulting $\Lambda_0$-values as a function of the flow time are shown on the right side of Fig.~\ref{fig:fixed_tau_lambda0} 
for different orders of perturbation theory used for  the zero flow time part of Eq.~\eqref{eq:pert_force_combined}.
We observe, that in the range $0.08\leq\tau_F/a^2_{20}\leq 0.18$, the flow time dependence of $\Lambda_0$ can be described by a constant within the error bars. 
Using a constant fit to extrapolate to the zero flow time limit, we find the values for $\Lambda_0$ at 
1-, 2-, and 3-loops with ultrasoft logs resummed 
to be \SI{0.297}{GeV}, \SI{0.264}{GeV}, and \SI{0.253}{GeV} respectively. 
Our result
is consistent with the value \SI{0.251(12)}{GeV} that was measured from 
the pure gauge static potential in Ref.~\cite{Brambilla:2010pp}.
We leave the full error analysis of these results to a future publication~\cite{futureforce}.

\acknowledgments{}
The lattice QCD calculations have been performed using the publicly available \href{https://web.physics.utah.edu/~detar/milc/milcv7.html}{MILC code}. The simulations were carried out on the computing facilities of the Computational Center for Particle and Astrophysics (C2PAP) in the project 'Calculation of finite T QCD correlators' (pr83pu) and of the SuperMUC cluster at the Leibniz-Rechenzentrum (LRZ) in the project 'The role of the charm-quark for the QCD coupling constant' (pn56bo). This research was funded by the Deutsche Forschungsgemeinschaft (DFG, German Research Foundation) cluster of excellence “ORIGINS” (\href{www.origins-cluster.de}{www.origins-cluster.de}) under Germany’s Excellence Strategy EXC-2094-390783311.

\bibliographystyle{JHEP}
\bibliography{bibliography.bib}

\end{document}